\documentclass[pre,showpacs,preprint]{revtex4}
\usepackage{amsmath}

\begin{document}
\title{Higher order correlations for fluctuations in the presence of 
fields}
\author{A. Boer}
\date{\today}
\email{boera@unitbv.ro}
\affiliation{Department of Physics, Transilvania University, B-dul 
Eroilor, R-2200, Bra\c {s}ov, Romania}

\author{S. Dumitru}
\email{s.dumitru@unitbv.ro}
\affiliation{Department of Physics, Transilvania University, B-dul 
Eroilor, R-2200, Bra\c {s}ov, Romania}

\begin{abstract}
The higher order moments of the fluctuations for the thermodynamical systems in the presence
of fields are investigated in the framework of a theoretical method. The method uses a generalized 
statistical ensemble consonant with the adequate expression for the internal energy.
The applications refer to the case of a system in magnetoquasistatic field. In the case of linear magnetic 
media one finds that  for the description of the magnetic induction fluctuations the Gaussian approximation 
is good enough. For nonlinear media the corresponding fluctuations are non-Gaussian, they having a 
non-null asymmetry. Additionally the respective fluctuations have characteristics of leptokurtic, 
mesokurtic and platykurtic type, depending on the value of the magnetic field strength comparatively
with a scaling factor of the magnetization curve.   
\end{abstract}

\maketitle

\section{Introduction}\label{sec:1}
In our previous work \cite{1} we have presented a phenomenological approach of the fluctuations for the generalized systems (i.e. thermodynamical systems in the presence of fields). In the respective work the fluctuations were evaluated only by second order numerical characteristics (correlations and moments). From a more general probabilistic perspective \cite{2,3,4,5,6} the fluctuations in physical systems must be evaluated also by means of higher order numerical characteristics (higher order correlations and moments). In the present paper we aim to present an approach for the evaluation of the higher order correlations for the alluded generalized systems. 

For our aim in Sec.\ref{sec:2} we will present the general considerations and relations regarding the approached problems. Next one in Sec.\ref{sec:3} we proceed to evaluations of higher order moments for linear magnetic systems (in the presence of magnetoquasistatic field), as well as for nonlinear magnetic media. The mentioned evaluations are finalized in some interesting relations able for comparations with adequate experimental results.

\section{Theoretical considerations}\label{sec:2}
\subsection{Statistical ensembles for generalized systems}
As in previous work \cite{1} (and above reminded) we consider a 
generalized system described by the set of extensive parameters 
$(U,X_{1},X_{2},\ldots,X_{n},Y_{1},Y_{2},\ldots,Y_{m})$, where $U$ 
denotes the generalized internal energy, $X_{i}\,(i=1,2,\ldots,n)$ 
signify the usual extensive parameters (for systems in absence of 
fields), while $Y_{j}\,(j=1,2,\ldots,m)$ mean the extensive parameters 
associated with the fields.

In the framework of fluctuation theory the thermodynamical quantities 
represent the mean (or expected) values of random variables. In the 
following we will note with $\overline{A}$ the mean value of the 
random quantity $A$. In such context we have $U=\overline{E}$, where 
$E$ denotes the energy regarded as random variable.    

We take the investigated system as a small part of a big, isolated 
ensemble. Generally speaking the probability density for the 
fluctuations of the quantities $E,\left\{X_{i}\right\}_{i=1}^{n}$ and 
$\left\{Y_{j}\right\}_{j=1}^{m}$ are of the form \cite{7,8,9}  
\begin{equation}\label{eq:1}
w=Z^{-1}\exp \left( -\beta E-\sum_{i=1}^{n}\alpha_{i}X_{i}
-\sum_{j=1}^{m}\gamma_{j}Y_{j} \right)
\end{equation}
where $Z$ is given by the normalization condition of the probability.
\begin{equation}\label{eq:2}
Z=\int\dotsi\int\prod_{i=1}^{n}\mathrm{d}X_{i}
\exp\left({-\alpha_{i}X_{i}}\right)
\prod_{j=1}^{m}\mathrm{d}Y_{j}\exp\left({-\gamma_{j}Y_{j}}\right) 
\int_{\Gamma}\mathrm{e}^{-\beta E}\mathrm{d}\Gamma
\end{equation}
$\mathrm{d}\Gamma$ represents the elementary volume in phase space. 

The statistical integral $Z$ is a function of the quantities $\beta$, 
$\left\{\alpha_{i}\right\}_{i=1}^{n}$ 
and $\left\{\gamma_{j}\right\}_{j=1}^{m}$. For the identification of 
the physical signification of the parameters
$\beta$, $\alpha_{i}$ and $\gamma_{j}$ we note 
\begin{eqnarray}\label{eq:3}
\mathrm{d}\left(\ln Z+\beta \overline{E} +\sum_{i=1}^{n}\alpha_{i} 
\overline{X_{i}}+\sum_{j=1}^{m}\gamma_{j}\overline{Y_{j}}
\right)=\nonumber\\
=\beta\mathrm{d}\overline{E}+\sum_{i=1}^{n}\alpha_{i}
\mathrm{d}\overline{X_{i}}
+\sum_{j=1}^{m}\gamma_{j}\mathrm{d}\overline{Y_{j}}
\end{eqnarray}
or
\begin{eqnarray}\label{eq:4}
\mathrm{d}\left(\ln Z+\beta U +\sum_{i=1}^{n}\alpha_{i} 
\overline{X_{i}}+\sum_{j=1}^{m}\gamma_{j}\overline{Y_{j}}
\right)=\nonumber\\
=\beta\mathrm{d}U+\sum_{i=1}^{n}\alpha_{i}
\mathrm{d}\overline{X_{i}}
+\sum_{j=1}^{m}\gamma_{j}\mathrm{d}\overline{Y_{j}}
\end{eqnarray}
where we take into account that in fact $U=\overline{E}$.

From the thermodynamics of the generalized systems it is known 
\cite{10,11} that
\begin{equation}\label{eq:5}
\mathrm{d}U=T\mathrm{d}S+\sum_{i=1}^{n}\widehat{\xi}_{i}\,
\mathrm{d}\overline{X_{i}}+\sum_{j=1}^{m}\psi_{j}\,\mathrm{d}
\overline{Y_{j}}
\end{equation}
where $S$, $T$ and $\xi_{i}$ denote respectively the entropy, 
temperature and the field dependent intensive 
parameters, while $\psi_{j}$ signify the supplementary parameters due 
to the presence of the fields.

From (\ref{eq:4}) and (\ref{eq:5}) one obtains:
\begin{eqnarray}\label{eq:6}
&&\mathrm{d}\left(\ln Z+\beta U+\sum_{i=1}^{n}\alpha_{i} 
\overline{X_{i}}+\sum_{j=1}^{m}\gamma_{j}\overline{Y_{j}}\right)=
\nonumber\\
&=&\beta T\mathrm{d}S+\sum_{i=1}^{n}\left(\alpha_{i}+\beta
\widehat{\xi}_{i}\right)\mathrm{d}\overline{X_{i}}+
\sum_{j=1}^{m}\left(\gamma_{j}+\beta\psi_{j}\right)
\mathrm{d}\overline{Y_{j}}
\end{eqnarray}
One observes that in this relation the left hand term is an exact 
differential. This fact impose the condition that the right hand term 
of the equation to be also an exact differential. From the mentioned 
condition it directly results that we must have the following 
relations:
\begin{equation}\label{eq:7}
\beta=\frac{1}{kT}
\end{equation}
\begin{equation}\label{eq:8}
\alpha_{i}=-\beta\widehat{\xi}_{i}=-\frac{\widehat{\xi}_{i}}{kT}
\end{equation}
\begin{equation}\label{eq:9}
\gamma_{j}=-\beta\psi_{j}=-\frac{\psi_{j}}{kT}
\end{equation}
In the above relations $k$ denotes the Boltzmann's constant.

We observe that the quantities $\left\{\alpha_{i}\right\}_{i=1}^{n}$ 
are functions of the field dependent intensive parameters.

\subsection{The evaluation of the higher order correlations and 
moments}
The mean values of the fluctuating quantities can be evaluated through 
the relations:
\begin{equation}\label{eq:10}
U=\overline{E}=-\left[\frac{\partial(\ln Z)}{\partial 
\beta}\right]_{\alpha_{i},\gamma_{j}},\qquad \quad
\begin{array}{ll}
i=1,2,\ldots ,n\\
j=1,2,\ldots ,m
\end{array}
\end{equation}
\begin{equation}\label{eq:11}
\overline{X_{i}}=-\left[\frac{\partial (\ln Z)}{\partial 
\alpha_{i}}\right]_{\beta,\alpha_{l},\gamma_{j}},\qquad \quad
l\neq i
\end{equation}
\begin{equation}\label{eq:12}
\overline{Y_{j}}=-\left[\frac{\partial (\ln Z)}{\partial 
\gamma_{j}}\right]_{\beta,\alpha_{i},\gamma_{l}},\qquad \quad
l\neq j
\end{equation}

The expressions of the type $\overline{\prod_{i}(\delta X_{i})^{r_{i}} 
\prod_{j}(\delta Y_{j})^{s_{j}}}$ are called higher order 
correlations. By using the statistical sum $Z$ as above introduced for 
some of the respective correlations one obtains the expressions:
\begin{equation}\label{eq:13}
\overline{\delta X_{a}\delta X_{b}}=\frac{\partial^2 (\ln Z)} 
{\partial\alpha_{a}\partial\alpha_{b}}=-\frac{\partial 
\overline{X_{b}}}{\partial\alpha_{a}}=kT\frac{\partial 
\overline{X_{b}}}{\partial\widehat{\xi}_{a}}\,,\qquad
a,b=1,2,\ldots ,n
\end{equation}
\begin{equation}\label{eq:14}
\overline{\delta Y_{a}\delta Y_{b}}=\frac{\partial^2 (\ln Z)} 
{\partial\gamma_{a}\partial\gamma_{b}}=-\frac{\partial 
\overline{Y_{b}}}{\partial\gamma_{a}}=kT\frac{\partial 
\overline{Y_{b}}}{\partial\psi_{a}}\,,\qquad
a,b=1,2,\ldots ,m
\end{equation}
\begin{equation}\label{eq:15}
\overline{\delta X_{a}\delta Y_{b}}=\frac{\partial^2 (\ln Z)} 
{\partial\alpha_{a}\partial\gamma_{b}}=-\frac{\partial 
\overline{Y_{b}}}{\partial\alpha_{a}}=kT\frac{\partial 
\overline{Y_{b}}}{\partial\widehat{\xi}_{a}}\,,\qquad
\begin{array}{ll}
a=1,2,\ldots ,n  \\
b=1,2,\ldots ,m
\end{array}
\end{equation}
\begin{equation}\label{eq:16}
\overline{\delta X_{a}\delta X_{b}\delta X_{c}}=
-\frac{\partial^3 (\ln Z)}{\partial 
\alpha_{a}\partial\alpha_{b}\partial\alpha_{c}}
=\frac{\partial^2 \overline{X_{c}}}{\partial \alpha_{a} 
\partial\alpha_{b}}=k^2 T^2 \frac{\partial^2 \overline{X_{c}}} 
{\partial\widehat{\xi}_{a}\partial\widehat{\xi}_{b}}\,,\qquad
a,b,c=1,2,\ldots n
\end{equation}
\begin{equation}\label{eq:17}
\overline{\delta Y_{a}\delta Y_{b}\delta Y_{c}}=
-\frac{\partial^3 (\ln Z)}{\partial 
\gamma_{a}\partial\gamma_{b}\partial\gamma_{c}}
=\frac{\partial^2 \overline{Y_{c}}}{\partial \gamma_{a} 
\partial\gamma_{b}}=k^2 T^2 \frac{\partial^2 \overline{X_{c}}} 
{\partial\psi_{a}\partial\psi_{b}}\,,\qquad
a,b,c=1,2,\ldots m
\end{equation}
\begin{equation}\label{eq:18}
\overline{\delta X_{a}\delta Y_{b}\delta Y_{c}}=
-\frac{\partial^3 (\ln Z)}{\partial 
\alpha_{a}\partial\gamma_{b}\partial\gamma_{c}}
=\frac{\partial^2 \overline{Y_{c}}}{\partial \alpha_{a} 
\partial\gamma_{b}}=k^2 T^2 \frac{\partial^2 \overline{Y_{c}}} 
{\partial\widehat{\xi}_{a}\partial\psi_{b}}\,,\qquad
\begin{array}{ll}
a=1,2,\ldots n  \\
b,c=1,2,\ldots ,m
\end{array}
\end{equation}

The formulae for the correlations of orders higher than 3 are 
generally more complicated. But the higher order moments can be 
obtained by means of the following recurrence formulae \cite{7}:
\begin{equation}\label{eq:19}
\overline{(\delta X_{a})^{n+1}}=-\frac{\partial}{\partial \alpha_{a}}
\overline{(\delta X_{a})^n}-n\overline{(\delta X_{a})^{n-1}}\;
\frac{\partial \overline{X_{a}}}{\partial \alpha_{a}}
\end{equation}
\begin{equation}\label{eq:20}
\overline{(\delta Y_{a})^{n+1}}=-\frac{\partial}{\partial \gamma_{a}}
\overline{(\delta Y_{a})^n}-n\overline{(\delta Y_{a})^{n-1}}\;
\frac{\partial \overline{Y_{a}}}{\partial \gamma_{a}}
\end{equation}

As examples we give here the expressions of the moments 
$\overline{(\delta X_{a})^4}$ and $\overline{(\delta Y_{a})^4}$.
\begin{eqnarray}\label{eq:21}
\overline{(\delta X_{a})^4}&=& -\frac{\partial^3 \overline{X_{a}}}
{\partial\alpha_{a}^3}+3\left(\frac{\partial\overline{X_{a}}} 
{\partial\alpha_{a}}\right)^2=  \nonumber \\
&=& \left( kT\frac{\partial}{\partial \widehat{\xi}_{a}} \right)^3
\overline{X_{a}}+3\left(kT\frac{\partial \overline{X_{a}}}
{\partial \widehat{\xi}_{a}}\right)^2
\end{eqnarray}
\begin{eqnarray}\label{eq:22}
\overline{(\delta Y_{a})^4}&=& -\frac{\partial^3 \overline{Y_{a}}}
{\partial\gamma_{a}^3}+3\left(\frac{\partial\overline{Y_{a}}} 
{\partial\gamma_{a}}\right)^2=  \nonumber \\
&=& \left( kT\frac{\partial}{\partial\psi_{a}} \right)^3
\overline{Y_{a}}+3\left(kT\frac{\partial \overline{Y_{a}}}
{\partial\psi_{a}}\right)^2
\end{eqnarray}

The fourth order moments are of interest for the evaluation of the so 
called excess coefficient:
\begin{equation}\label{eq:23}
C_{E}=\frac{\overline{(\delta X_{a})^4}}
{\left[\overline{(\delta X_{a})^2}\right]^2}-3
\end{equation}
which is an indicator of the deviation from the gaussian distribution 
\cite{2}.

\section{Higher order moments for the systems situated in a \\ 
magnetoquasistatic field}\label{sec:3}

\subsection{Linear magnetic media}
Let us consider a uniformly magnetized continuous media, situated in a 
magnetoquasistatic field. 
The system is characterized by the extensive parameters 
$(U,V,N,\mathbf{B})$, where $V$, $N$ and $\mathbf{B}$ 
denote respectively the volume, particle number and magnetic 
induction. In the case of linear magnetic systems the differential of 
the internal energy is given by \cite{10,11}
\begin{equation}\label{eq:24}
dU=T\mathrm{d}S-\widehat{p}\,\mathrm{d}V+\widehat{\zeta}\mathrm{d}N+
V\mathbf{H}\cdot \mathrm{d}\mathbf{B}
\end{equation}

For the sake of brevity we omitted the mean symbol from above the 
parameters $V,N,\mathbf{B}$, but we will
take into account that in fact in relation (\ref{eq:24}) and in the 
expressions of the moments appear the mean 
values of the respective quantities. 

The relations (\ref{eq:13}) and (\ref{eq:16}) show that the moments 
associated to the usual thermodynamic 
quantities, i.e. volume $V$ or particle number $N$, are functions of 
the parameters $\widehat{\xi}_{i}$, which depend on the
fields constraints. 

For example, in the case $\mathbf{B}=\text{const.}$ for the volume $V$ 
one obtains:
\begin{equation}\label{eq:25}
\overline{(\delta V)^2}=-kT\left(\frac{\partial V}
{\partial\widehat{p}}\right)_{T,\,\widehat{\zeta},\,\mathbf{B}}
\end{equation}
\begin{equation}\label{eq:26}
\overline{(\delta V)^3}=k^{2}T^{2}\left(\frac{\partial^2 V}
{\partial\widehat{p}\,^{2}}\right)_{T,\,\widehat{\zeta},\,\mathbf{B}}
\end{equation}
where \cite{10}
\begin{equation}\label{eq:27}
\widehat{p}\,(\mathbf{B}=\text{const.})=p_{\mathbf{B},N}=
p-\frac{1}{2}\mathbf{H}\cdot\mathbf{B}-\frac{1}{2}H^2\rho
\frac{\partial\mu}{\partial\rho}
\end{equation}
$\mathbf{H}$ signify the magnetic field strength, $\mu$ is the 
magnetic permeability and $\rho=\frac{N}{V}$.

By using the properties of the Jacobians the relation (\ref{eq:25}) 
can be transformed as follows:
\begin{eqnarray}\label{eq:28}
\overline{(\delta V)^2}&=&-kT\,\frac{\partial (V,\widehat{\zeta},
T,\mathbf{B})}{\partial (\widehat{p},\widehat{\zeta},T,\mathbf{B})}
=-kT\,\frac{\partial (V,\widehat{\zeta},T,\mathbf{B})}
{\partial (V,N,T,\mathbf{B})}\cdot \frac{\partial (V,N,T,\mathbf{B})}
{\partial (\widehat{p},\widehat{\zeta},T,\mathbf{B})}=\nonumber\\
&=&-kT\,\frac{\left(\frac{\partial\widehat{\zeta}}{\partial N}\right)
_{T,V,\mathbf{B}}}{\left(\frac{\partial\widehat{p}}{\partial V}\right)
_{T,N,\mathbf{B}}\left(\frac{\partial\widehat{\zeta}}{\partial 
N}\right)_{T,V,\mathbf{B}}+\left(\frac{\partial\widehat{\zeta}}
{\partial V}\right)^2_{T,N,\mathbf{B}}}
\end{eqnarray}
In the above relation we take into account the condition 
$-\left(\frac{\partial\widehat{p}}{\partial N}\right)_{T,V,\mathbf{B}} 
=\left(\frac{\partial\widehat{\zeta}} 
{\partial V}\right)_{T,N,\mathbf{B}}$.
Here is the place to point out that the result (\ref{eq:28}) is the 
same with the one obtained \cite{1} within
the Gaussian approximation.

Now let us evaluate the second and third order parameters of 
fluctuations for the magnetic induction $\mathbf{B}$.
For simplicity we consider that volume and particle number are 
constant. By using the relations (\ref{eq:14}), (\ref{eq:17}) and 
(\ref{eq:22}) we find:
\begin{equation}\label{eq:29}
\overline{(\delta B)^2}=\frac{kT}{V}\left(\frac{\partial B}
{\partial H}\right)_{T,V,N}=\frac{kT\mu}{V}
\end{equation}
\begin{equation}\label{eq:30}
\overline{(\delta B)^3}=\left(\frac{kT}{V}\right)^2\left(\frac{\partial^2 B}
{\partial H^2}\right)_{T,V,N}=\left(\frac{kT}{V}\right)^2
\frac{\partial\mu}{\partial H}=0
\end{equation}

The result (\ref{eq:29}) is identical with the one obtained \cite{1} 
within the Gaussian approximation. We remark that in the case of 
linear magnetic media $\overline{(\delta B)^3}$ is null, because $\mu$ 
is independent on $\mathbf{H}$. Additionally in this case the excess
coefficient (\ref{eq:23}) is also null. These facts show that in the 
alluded case the Gaussian approximation is sufficient for a quantitative
description of the fluctuations for $B$.  

\subsection{Nonlinear magnetic media}
In the case of nonlinear magnetic media $\mu$ depends on $\mathbf{H}$. 
Therefore the evaluation of the moments of orders higher than 2 become 
necessary.

We approach such a case under the constraints when $V=\text{const.}$ 
and $N=\text{const}$. Then for the 
internal energy $U$ we have:
\begin{equation*}
\mathrm{d}U=T\mathrm{d}S+V\mathbf{H}\cdot\mathrm{d}\mathbf{B}
\end{equation*}

For the moments of 2, 3 and 4 order of $\mathbf{B}$ we obtain:  
\begin{equation}\label{eq:31}
\overline{(\delta B)^2}=\frac{kT}{V}\left(\frac{\partial B}
{\partial H}\right)_{T,V,N}
\end{equation}
\begin{equation}\label{eq:32}
\overline{(\delta B)^3}=\left(\frac{kT}{V}\right)^2 
\left(\frac{\partial^{2}B}{\partial H^2}\right)_{T,V,N}
\end{equation}
\begin{eqnarray}\label{eq:33}
\overline{\left(\delta 
B\right)^4}&=&\left(\frac{kT}{V}\right)^3\left(\frac{\partial^{3}B}
{\partial H^3}\right)_{T,V,N}+
3\left[\frac{kT}{V}\left(\frac{\partial B}
{\partial H}\right)_{T,V,N}\right]^2=  \nonumber \\
&=&\left(\frac{kT}{V}\right)^3\left(\frac{\partial^{3}B}
{\partial H^3}\right)_{T,V,N}+3\left[\overline{(\delta B)^2}\right]^2
\end{eqnarray}

For finding the explicit expressions of $\overline{(\delta B)^2}$, 
$\overline{(\delta B)^3}$ and $\overline{(\delta B)^4}$ it is 
necessary to know the expression of the function $B=B(H)$. 
The most known such function is given by the Langevin equation:
\begin{equation}\label{eq:34}
B=\mu_{0}M_{s}\left(\coth{a}-\frac{1}{a}\right)+\mu_{0}H
\end{equation}
where 
\begin{equation}\label{eq:35}
a=\frac{\mu_{0}mH}{kT}
\end{equation}
$M_{s}$ represents the saturation magnetization, $\mu_{0}$ is the 
vacuum permeability and $m$ signify 
the magnetic moment of an individual molecule.

By means of some simple mathematical operations one finds:
\begin{equation}\label{eq:36}
\overline{(\delta B)^2}=\frac{kT\mu_{0}}{V}\left[
\frac{\mu_{0}mM_{s}}{kT}\left(\frac{1}{a^2}-\frac{1}
{\sinh^2{a}}\right)+1\right]
\end{equation}
\begin{equation}\label{eq:37}
\overline{(\delta B)^3}=\frac{2\mu_{0}^{3}m^{2}M_{s}}{V^2}
\left(\frac{\cosh a}{\sinh^3{a}}-\frac{1}{a^3}\right)
\end{equation}
\begin{eqnarray}\label{eq:38}
\overline{(\delta B)^4}&=&\frac{2\mu_{0}^{4}m^{3}M_{s}}{V^3}
\left(\frac{3}{a^4}+\frac{\sinh^{2}a-3\cosh^{2}a}{\sinh^{4}a}\right)+
\nonumber \\
&&+3\left\{\frac{kT\mu_{0}}{V}\left[\frac{\mu_{0}mM_{s}}
{kT}\left(\frac{1}{a^2}-\frac{1}{\sinh^{2}a}\right)+1\right]\right\}^2
\end{eqnarray}

One of the most used dependence of $B$ from $H$
is given \cite{12} by the function 
\begin{equation}\label{eq:39}
B=\mu_{0}M_{s}\left[1-\exp\left(-\frac{H^2}{2\sigma^2}\right)\right]+
\mu_{0}H
\end{equation}
where $\sigma$ is a scaling factor. 

The mentioned function imply a maximum for the differential permeability
$\mu_{d}={\mathrm{d}B}/{\mathrm{d}H}$ at the value $H=\sigma$.

By using the general formulas for the 2nd, 3rd and 4th order moments of the 
random variable $B$ one obtains: 
\begin{equation}\label{eq:40}
\overline{\left(\delta B\right)^2}=
\frac{\mu_{0}kT}{V}\left\{\frac{M_{s}H}{\sigma^2}
\exp\left(-\frac{H^2}{2\sigma^2}\right)+1\right\}
\end{equation}
\begin{equation}\label{eq:41}
\overline{\left(\delta B\right)^3}=
\left(\frac{kT}{V}\right)^2\frac{\mu_{0}M_{s}}{\sigma^2}
\left(1-\frac{H^2}{\sigma^2}\right)
\exp\left(-\frac{H^2}{2\sigma^2}\right)
\end{equation}
\begin{eqnarray}\label{eq:42}
\overline{\left(\delta B\right)^4}&=&
\left(\frac{kT}{V}\right)^3\frac{\mu_{0}M_{s}H}{\sigma^4}
\left(\frac{H^2}{\sigma^2}-3\right)
\exp\left({-\frac{H^2}{2\sigma^2}}\right)+ \nonumber \\
&&+3\left(\frac{\mu_0 kT}{V}\right)^2
\left\{\frac{M_{s}H}{\sigma^2}
\exp\left(-\frac{H^2}{2\sigma^2}\right)+1\right\}^2
\end{eqnarray}

In the end we wish to note the following observations:
\begin{enumerate}
\item $\overline{(\delta B)^3}$ change its sign in the point 
$H=\sigma$, where the differential permeability \\
$\mu_{d}=\frac{\mathrm{d}B}{\mathrm{d}H}$ takes its maximal value 
[this means the inflection point of 
the function $B=B(H)$]. For $H<\sigma$, the moment $\overline{(\delta B)^3}$ is 
positive while for $H>\sigma$ it is negative.
\item The fourth order moment $\overline{(\delta B)^4}$ give 
informations about the deviation from the 
Gaussian approximation. This because it is implied in the so called 
excess coefficient (\ref{eq:23}).
In the here discussed case we have:
\begin{equation}\label{eq:43}
C_E=\frac{kT}{V}\frac{M_{s}H}{\mu_{0}\sigma^4}
\frac{\left(\frac{H^2}{\sigma^2}-3\right)
\exp\left(-\frac{H^2}{2\sigma^2}\right)}
{\left\{\frac{M_{s}H}{\sigma^2}
\exp\left(-\frac{H^2}{2\sigma^2}\right)+1\right\}^2}
\end{equation}
In probabilistic terminology \cite{13} the distribution of a random variable is
called leptokurtic, mesokurtic or platykurtic as the excess coefficient 
$C_E$ satisfy the conditions $C_E>0$, $C_E=0$ and $C_E<0$ respectively. Then by 
taking into account the expression (\ref{eq:43}) of $C_E$ one can say that for the
here studied situation the fluctuations of the magnetic induction $B$ are leptokurtic, 
mesokurtic and platykurtic as the magnetic field strength $H$ satisfy the 
conditions \mbox{$H>\sqrt{3}\,\sigma$}, \mbox{$H=\sqrt{3}\,\sigma$} and 
\mbox{$H<\sqrt{3}\,\sigma$} respectively.  
\end{enumerate}

\section{Summary and conclusions}
\begin{enumerate}
\item We investigated the higher order moments of the fluctuations for complex 
thermodynamic systems (i.e. systems considered in the presence of fields). Our approach 
uses a generalized statistical ensemble. We considered the case when the energy $E$, 
the usual extensive parameters $\left\{X_i\right\}_{i=1}^{n}$ 
and the field parameters $\left\{Y_j\right\}_{j=1}^{m}$ are fluctuating random variables.
We find that the higher order moments of fluctuations depend on the field constraints.
\item The general results from the Sec.\ref{sec:2} were particularized for the case of 
a system situated in a magnetoquasistatic field. If the magnetic characteristics of such
system are linear the third order moment of the magnetic induction and the excess 
coefficient are null. Therefore the description of the fluctuations for the magnetic
induction can be done in the framework of Gaussian approximation.
\item For nonlinear magnetic media \mbox{$\overline{\left(\delta B\right)^3}\neq 0$} and consequently
the fluctuations of $B$ deviate from the normal distribution. Additionally the respective 
deviation are characterized by the various value of the excess coefficient $C_E$ given by the
formula (\ref{eq:43}). From the respective formula it results that the fluctuations of $B$ can
be leptokurtic (for $H>\sqrt{3}\,\sigma$), mesokurtic (for $H=\sqrt{3}\,\sigma$) and 
respectively platykurtic (for $H<\sqrt{3}\,\sigma$) for the cases in which the function 
$B=B(H)$ is given by the relation (\ref{eq:39}).    
\end{enumerate}

\begin{acknowledgments}
Our researches reported both here and in a previous paper were stimulated by
the publications of Professor Y. Zimmels. We express our gratitude to Prof. Zimmels
for putting at our disposal a lot of his publications. We mention also that the
present work takes advantage of some facilities of a grant from the Romanian
Ministry of Education and Research.
  
\end{acknowledgments}

\end{document}